\documentstyle[12pt]{article}
\let\nopictures=N
\newcommand{\be}{\begin{equation}}
\newcommand{\ee}{\end{equation}}
\newcommand{\bea}{\begin{eqnarray}}
\newcommand{\eea}{\end{eqnarray}}
\newcommand{\nen}{\nonumber \\ \relax}
\renewcommand{\theequation}{\thesubsection.\arabic{equation}}
\newcommand{\forcepar}{{\hskip 10pt\vskip -15pt}}
\newfont{\headfont}{cmbx10 scaled 1440}
\newfont{\namefont}{cmr10}
\newfont{\initialfont}{cmr10 scaled 1200}
\newfont{\addfont}{cmti7 scaled 1440}
\newfont{\boldmathfont}{cmbx10}
\newfont{\figfont}{cmr7 scaled 1200}

\newcommand{\seq}{\ =\ }
\newcommand{\pls}{\ +\ }
\newcommand{\mi}{\ -\ }

\newcommand{\half}{\frac{1}{2}}

\newcommand{\g}{\gamma}

\newcommand{\ca}{{\cal A}}
\newcommand{\cb}{{\cal B}}
\newcommand{\cc}{{\cal C}}
\newcommand{\cd}{{\cal D}}

\newcommand{\cg}{{\cal G}}

\newcommand{\cl}{{\cal L}}
\newcommand{\cm}{{\cal M}}
\newcommand{\cn}{{\cal N}}
\newcommand{\co}{{\cal O}}

\newcommand{\IR}{{I \kern -0.4em R}}
\newcommand{\IC}{{I \kern -0.65em C}}
\newcommand{\prd}[1]{{\it Phys. Rev.} {\bf D#1}}

\newcommand{\np}[1]{{\it Nucl. Phys.} {\bf B#1}}
\newcommand{\cmp}[1]{{\it Commun. Math. Phys.} {\bf #1}}
\newcommand{\ijmp}[1]{{\it Intl. J. Mod. Phys.} {\bf A#1}}

\newcommand{\cqg}[1]{{\it Class. Quan. Grav.} {\bf #1}}

\newcommand{\prl}[1]{{\it Phys. Rev. Lett.} {\bf #1}}

\newcommand{\pl}[1]{{\it Phys. Lett.} {\bf #1B}}

\newfont{\headfontb}{cmbx10 scaled 1728}
\newcommand{\Abun}{{\bf A}}
\newcommand{\dbun}{{\bf d}}
\newcommand{\Fbun}{{\bf F}}

\newcommand{\ddelta}{ \delta}

\newcommand{\dual}{^\star\!}
\def\ev#1{\left\langle#1\right\rangle}

\def\bft#1{BF-gauge theory#1}
\def\bfts#1{BF-gauge theories#1}
\def\sbft#1{super-BF gauge theory#1}
\def\sbfts#1{super-BF gauge theories#1}
\begin{document}
\begin{titlepage}
\renewcommand{\thefootnote}{\fnsymbol{footnote}}
\begin{center}
{\headfontb Quantum Gravity\\ and Equivariant
Cohomology}\footnote{This work is supported in part by funds
provided by the
U. S. Department of Energy (D.O.E.) under cooperative agreement
\#DE-FC02-94ER40818.}

\end{center}
\vskip 0.3truein
\begin{center}
{
{\Large R}{OGER}
		    {\Large B}{ROOKS\footnote{E-mail: rog@ctpup.mit.edu}}
{ AND}
{\Large G}{ILAD}
		   {\Large L}{IFSCHYTZ\footnote{E-mail: gil1@irene.mit.edu}} }
\end{center}
\begin{center}
{\addfont{Center for Theoretical Physics,}}\\
{\addfont{Laboratory
for Nuclear Science}}\\
{\addfont{and Department of Physics,}}\\
{\addfont{Massachusetts Institute of Technology}}\\
{\addfont{Cambridge, Massachusetts 02139 U.S.A.}}
\end{center}
\vskip 0.5truein
\begin{abstract}
A procedure for obtaining correlation function densities and
wavefunctionals for quantum gravity from the Donaldson polynomial
invariants of topological quantum field theories, is given.  We
illustrate how our procedure may be applied to three and four
dimensional quantum gravity.  Detailed expressions, derived from
\sbft{}, are given in the three dimensional case.  A procedure for
normalizing these wavefunctionals is proposed.
\vskip 0.5truein
\leftline{CTP \# 2340 \hfill July 1994}
\smallskip
\leftline{hep-th/9407177}
\end{abstract}
\vskip 0.5truein
\end{titlepage}

\setcounter{footnote}{0}

\section{Introduction}\medskip\forcepar
Topological invariants on a
manifold are a subset of diffeomorphism invariants.  Thus we expect
that elements of the set of topological invariants should be a subset
of the quantum gravity observables.  Additionally, it is generally
believed that observables, which are elements of the BRST complex,
may be used to construct vertex operators or wavefunctionals for the
theory.  Consequently, should we succeed in constructing observables
for quantum gravity, we might also be able to construct
wavefunctionals.  These statements form the nexus for the present
work.  The puzzle is how to find representations of topological
invariants in quantum gravity theories in sufficient generality so as
not to explicitly exploit the topological nature of low-dimensional
gravitational theories.  In this paper, we will give a formal
procedure for constructing operators which have the interpretation as
the densities of correlation functions of observables and which lead
to wavefunctionals, in this fashion.

Loop observables, which are constructed from Wilson loops, have been
proposed \cite{RovSmol, Smol} for four dimensional canonical gravity
in the Ashtekar formalism \cite{Ash} via the loop representation
\cite{(AshLoopRep)}.  In this way, observables which measure the
areas of surfaces and volumes of regions have been constructed
\cite{Smol}.  These are intricate constructions and we wonder if they
may be placed in a different context via appealing to the geometry of
the space of solutions to the constraints.  From observables, we
expect to be able to find states, and, perhaps, their
wavefunctionals.  Put into focus, our quest for a geometrical
interpretation for general quantum gravity is a hope that we may be
able to exploit the geometry to directly construct wavefunctionals.
This is not to mean that we are diminishing the importance of observables.

Indeed, the geometry which underlies gauge field theories suggests
another way of representing wavefunctionals; this will be the focal point
of our exploration in this work.  In particular, as both three
dimensional gravity \cite{Wit(CSGrav),Ash} and the \sbft{}
\cite{TFT_rev} of flat $SO(2,1)$ or $SO(3)$ connections (in which the
geometry of the space of connections is explicit) share the same
moduli space, these theories are natural choices for experimentation
on this idea.  We will find an interesting relation between the
polynomial topological invariants of three dimensional flat
connection bundles, which are the analogs of Donaldson's invariants
\cite{Don} for self-dual connections in four dimensional Yang-Mills
gauge theory, and correlation densities of three dimensional quantum
gravity.  This does not mean
that we will find correlation densities of new observables.  We
expect that the ones we will
obtain may be decomposed in terms of Wilson loops.  Further pursuit
of our ideas then lead us to expressions for canonical and
Hartle-Hawking wavefunctionals which satisfy the constraints of three
dimension gravity.  By exploiting previous work on four dimensional
topological gravity, we are also able to sketch how our approach
works in this physical dimension.  Due to the fact that much more is
known about the associated three dimensional topological quantum
field theories (TQFT's) \cite{TFT_rev} than four dimensional
topological gravity, we are presently unable to give expressions
which are as detailed as those for three dimensional quantum gravity.
We should point out that while the correlation densities and wavefunctionals
which may be constructed via our approach for three dimensional
gravity are likely to span the full space of such quantities, we do
not expect this to be the case for four dimensional gravity.  The
reason is simply that the phase space of the former theory and of
TQFT's is finite dimensional while that of the latter is not.

Our work is relatively formal as our objective is to establish an
approach to solving some of these long standing problems of quantum
gravity.  In particular, we give expressions in terms of path
integrals which, in principle, may be computed exactly.  These path
integrals appear as those of topological quantum field theories which
are strongly believed to be, at worst, renormalizable \cite{TFT_rev}.
This allows us to make use of BRST analysis techniques in order to
establish our results.  A related approach for the computation of
scattering amplitudes in string theory was undertaken by one of us in
ref. \cite{Bro}.

Commencing, we establish the framework of our approach while
attempting to be as general as possible, in the next section.
Implementation of the approach is carried out for three dimensional
\bft{}, in general, and 3D quantum gravity, in particular, in section
\ref{Sec_3D}.  Expressions for correlation densities are given in sub-section
\ref{SSec_Obs} while wavefunctionals may be found in sub-section
\ref{SS_BFWfcn}.  The four dimensional case is sketched in section
\ref{Sec_4D}.  Our conclusions may be found following that section.
In addition, appendices summarizing \bfts{} and \sbfts{} are
given.
In appendix \ref{A_Spec} we suggest the possible existence of  polynomial
invariants  in pure three quantum gravity, before applying our approach.
Our global notations are given in appendix \ref{A_Nota}.

\vskip 0.5truein
\section{The Heuristic Construction}\label{Sec_Idea}
\setcounter{equation}{0} \medskip \forcepar
As was discussed in the
introduction, our approach is to first find correlation densitites and then
extract the wavefunctionals from them.  Thus, in this section, we
first concentrate on our general approach to obtaining the
correlation densities.  Then, we will discuss how to obtain the
wavefunctionals
from them, at the end of this section.
\bigskip
\setcounter{equation}{0}
\subsection{Correlation Densities}\medskip\forcepar
Given a field theory, one is interested in its physical states and the
observables; {\it  i.e},  functionals and functions of the fields which obey
the constraint of the theory.  One reason why observables are
important is that from them physical correlation functions can be
constructed. However, it is not necessary to find observables in order
to construct physical correlation functions of the fields.  As an
example, given a function, $\hat\co$, of the fields, we will only
demand that the vacuum expectation value
 $\frac{\delta\ev{\hat\co}}{\delta g_{\mu\nu}}$
vanishes, where
$g_{\mu\nu}$ is some background metric.  This allows for
$\frac{\delta\hat\co}{\delta g_{\mu\nu}}\neq 0$.
 The ${\hat \co}_i$ we will construct will
have the property that generally
$\frac{\delta\ev{\prod_{i=1}^n{\hat\co}}}{\delta g_{\mu\nu}}\neq0$ for
$n\geq2$.  Thus they are really physical correlation function densities.
In this section, we will describe how we can use TQFT's in order to
construct, a set of the $\hat\co$'s for a general field theory (GFT).
Our focus will be on quantum gravity for which topological
observables are of interest.

Take a GFT for fields, $X$, which are sections of a bundle over a
manifold, $M$, and whose
space of physical fields is called $\cn$.
Construct \cite{Cobi} a TQFT which describes the geometry of a
subspace of $\cn$, which
we call $\cm$ (the dimension of $\cm$ is  finite). In this way,  we have
projected
the GFT onto the TQFT.
Expectation values of observables in the TQFT
(which we generally know how to write), are topological invariants of
$M$. Now if the TQFT has the constraints, $\cg$,  of the GFT
as a subset of it's own constraints then  we can construct physical
correlation functions and wavefunctionals of the GFT,
with the use of the TQFT.
We now describe two different ways of doing this.

First, suppose we are given a particular GFT for fields $X$ and are able to
construct a TQFT with fields $X$ and $Y$.  Let us require that this
TQFT has the same Lagrangian  as the GFT plus additional terms which
are also invariant under the local symmetries of the GFT\footnote{$Y$
may be thought of as the supersymmetric partners of $X$ and the
additional terms in the action as supersymmetric completion.}.
Furthermore, we require that
a subalgebra of the constraints of our TQFT is isomorphic to the
constraint algebra of the GFT.  In particular, the action of this
subset on the $X$, in the TQFT is the same as the action of the
GFT's constraints on $X$.  As an example, take the GFT to be \bft{}
and the TQFT to be \sbft.

Now take a set of observables
in the TQFT and {\it almost} compute their correlation function.
By this we mean the following.  Integrate the
path integral
over all the fields that are present in the TQFT but not in the GFT; that
is, over $Y$.  We then get an expression (which is
typically non-local), $\hat\co$,
in terms of the fields $X$. The
expectation value of $\hat\co$, in the GFT,
is a topological invariant of $M$. More precisely,
\be
\ev{\hat\co}_{GFT}\seq\int [dX] e^{-S_{GFT}}\hat\co (X) \seq \int
[dX][dY] e^{-S_{TQFT}}
\co(X,Y)\ \ ,\ee
where $\co$ is a product of observables in the TQFT, $S_{GFT}$ is the action
of the GFT and $\hat\co$ is a gauge invariant, non-local expression in terms of
the
original fields.
Really what we are doing is taking the original theory and coupling special
``matter'' to it, and using the matter part to construct physical
correlation functions.
However, using TQFT's has additional rewards.  First, these
expressions are computable as the theories are, at worst, renormalizable.
Second, we will see that we will be able
to write expressions without integrating over the entire spacetime
manifold, which satisfy the constraints of quantum gravity.

Although it is tempting to call the correlation densities
observables, this can only be done with qualification as they do not
have one of the important properties we associate with observables.
That is, generally the product of two or more of them is not an observables in
the sense that this product's vacuum expectation value will not be
diffeomorphism invariant.

To place the above arguments in a geometrical setting, let us look at the
geometry of the space of connections \cite{AtiySing}.  This argument
applies, in principle, to any gauge theory built from a Yang-Mills fields
space. Let $P(M,G)$ be a $G$-bundle over the spacetime manifold, $M$,
and $\ca$ be the  space of its connections.  Forms on the space
$P\times\ca$, $\Omega^{(p,q)}(P\times\ca)$, will be bi-graded
inheriting degrees $p$ from $M$ and $q$ from $\ca$.  A connection
$\Abun\equiv A+c$ may be introduced on the bundle $P\times\ca$ along
with an exterior derivative $\dbun\equiv d+Q$ where $d(Q)$ is the
exterior derivative on $M(\ca)$.  The object, $c$ is the ghost field
of the Yang-Mills gauge theory.  The total form degree of $\Abun$ is
one and is given by  the sum of the degree on $M$ and ghost number.
The curvature of the connection $\Abun$ is $\Fbun= \dbun \Abun +\half
[\Abun,\Abun] = F+\psi+\phi$, where the $(2,0)$ form $F = dA+A\wedge
A$ is the usual curvature of $P$, $\psi=QA+ d_Ac$ is a $(1,1)$ form
and $\phi=Qc+\half [c,c]$ is a $(0,2)$ form.  Gauge invariant and
metric independent operators may be constructed out of these objects.
They are the Donaldson invariants written in a field theoretic
language.  Thus we will be attempting to recover these geometrical objects
which already exist, but are hidden, in physical gauge theories.

\bigskip
\subsection{Wavefunctionals}\medskip\forcepar
\setcounter{equation}{0}
We can also obtain wavefunctionals of the GFT's fields, $X$,
which satisfy the  latter theories constraints.  A general method
will be described first, then another prescription which we will
later see
works for three dimensional gravity, but which is not guaranteed to
work in general, will be given.  In the following we will use the
term geometrical sector to refer to those fields which are realized
as the curvature components for the geometry of the universal bundle
over $X$.  For example, these would be $(A,\psi,\phi)$ in a theory
defined over a Yang-Mills field space.

First, take the TQFT to be
defined over a spacetime manifold $ M $
with a boundary, $\partial M$, which is homeomorphic to the surface, $\Sigma$,
we wish to quantize the GFT on.  As for the GFT, let the phase space
of the TQFT be even-dimensional.   Note
that $M$ need not be diffeomorphic to $\Sigma\times\IR$.
Form the correlation function of a set of observables in this TQFT.
Choose a polarization and
functionally integrate over the  $X$ and $Y$ sets of fields
in the TQFT with boundary conditions on $\Sigma$.  Then, the correlation
function will
yield a functional of the boundary values of half of the Cauchy data for the
$X$ fields, call that set
$ X|_{\Sigma}$,  and half of the $Y$ fields, call that set $Y|_{\Sigma}$.
By construction this is a Hartle-Hawking wavefunctional for the TQFT
which is guaranteed to be computable since, at worst, TQFT's are
renormalizable:
\be
\Psi[X|_{\Sigma},Y|_{\Sigma}]\seq\int [dX][dY] e^{-S_{TQFT}} \co(X,Y)\
\ .\ee
Here $S_{TQFT}$ is the TQFT action on the manifold with boundary,
$ \Sigma$. The wavefunctional, $\Psi[X|_{\Sigma},Y|_{\Sigma}]$ is
diffeomorphism
invariant due to the properties of TQFT's.  For the particular TQFT,
any fields which appear in $\Psi[X|_{\Sigma},Y|_{\Sigma}]$ and which
are not in the geometrical sector, should be integrated out.  Then
all the $Y$ fields which remain in $\Psi[X|_{\Sigma},Y|_{\Sigma}]$
may be replaced by non-local expressions involving $X$ and
$\frac{\partial X}{\partial m}$.  This is an idiosyncrasy of TQFT's.
Then since $\frac{\partial X}{\partial m}$ is a function of $X$, we
obtain
\be
\Psi[X|_\Sigma,Y|_\Sigma]~ \Longrightarrow~ \Psi[X|_\Sigma]\ \ .\ee
In practice, we find that those $Y|_\Sigma$ fields which appear in
$\Psi[X|_\Sigma,Y|_\Sigma]$ are Grassmann-odd and the projection to
$\Psi[X|_\Sigma]$ stated above is performed by first choosing a basis
for $T^*\cm$, expanding those $Y|_\Sigma$ in this basis and then
expanding $\Psi[X|_\Sigma,Y|_\Sigma]$ as a superfield whose
components are wavefunctionals, $\Psi[X|_\Sigma]$.

This approach leads us to the following ansatz for a normalization
procedure which stems from the axiomatic approach \cite{Atiyah} to TQFT's.
Given two wavefunctionals, $\Psi_1$ and $\Psi_2$, defined on diffeomorphic
boundaries, $\partial M_1$ and $\partial M_2$, we
might try defining the inner product
by gluing the two manifolds together. This will result in a path integral of
some observable of the TQFT defined on the glued manifold.
As these expressions
are finite this gives a possible normalization procedure. We differ the
exact construction to future work \cite{BroLif}.

A second approach to constructing the wavefunctionals stems from the
observation that, in the above, we took the wavefunctionals of the
TQFT and projected onto the $X$ subspace to obtain the
wavefunctionals of the GFT.  Thus it is suggestive to simply
construct the wavefunctionals of the TQFT by any means possible and
then apply the projection.  Thus we need not restrict ourselves to
Hartle-Hawking wavefunctionals but might also consider those obtained
by directly analyzing the constraints of the canonically quantized TQFT.

Now let us specialize to a certain set of GFT's.  For
certain theories, such as three dimensional gravity, we may construct
such wavefunctionals by building TQFT's which are defined in a
background which solves the constraints of the GFT.  We will call such
TQFT's, servant theories. That is, in the GFT, we
solve the constraints first and then quantize.  The quantization then
demands that we find wavefunctionals which have support only on the
constraints' solutions.  Realizing this, we construct correlation
functions in a servant TQFT which is defined over a certain background. As we
will see in the next section, this works when the servant TFT is of the
Schwarz \cite{Sch} type.   As the servant TFT's must be topological, this
approach restricts the background; {\it i. e.}, those $X|_\Sigma$ which
solve the constraints, to be non-propagating fields or global data.
Thus we expect that this approach will only work for certain sectors
of four dimensional gravity.

Having given a cursory discussion of our procedures for obtaining
observables and wavefunctionals, let us now turn to some specific
applications.   Three dimensional quantum gravity and \bfts{}, in
general, are first.

\vskip 0.5truein
\section{Application to 3D BF Gauge
Theories}\label{Sec_3D}
\setcounter{equation}{0}
\medskip \forcepar
As \bfts{} are TFT's, they are the logical choice for the first
application of the ideas discussed in the previous section.  Although
our analysis below may be carried out in arbitrary dimensions, we
will focus on 2+1 dimensional manifolds.  In this dimension, \bfts{}
are of more than a passing interest; as with gauge group $G=SO(2,1)$,
they are known \cite{Wit(CSGrav),Ash,AshHus_etal} to be theories of
quantum gravity.  In subsection \ref{SSec_Obs}, we will study the
construction of correlation densities in the covariant quantization of \bfts{}
based on the geometry of the universal bundle.  Then in subsection
\ref{SS_BFWfcn} we will give formal expressions for canonical and
Hartle-Hawking wavefunctionals of \bfts{} again based on the geometry
of the universal bundle.  Where appropriate, we will make allusions
to three dimensional quantum gravity

Before proceeding we would like to be further explain  the
rationale for choosing \bfts{} (see appendix \ref{A_BF}) as a first
application of our constructions.  There are cohomological field
theories (or TQFT's), called \sbfts, which share the same moduli
space.  As quantum field theories, they are very closely related
\cite{BroGat} and the manifest appearance of the geometry of the
constraint space of \bfts{} in the \sbfts{} will be most useful.
These two facets make the construction of observables and
wavefunctionals for \bfts{} from \sbfts{} highly suggestive and, as
we will find, possible.

\bigskip \subsection{Pulling Back $H^*(\cm)$ to
\bfts{}}\label{SSec_Obs} \medskip \forcepar
Define $\cn$ to be the
restriction of $\ca$ to flat connections: $\cn\leftrightarrow
\ca\big|_{F=0}$ and $\cm=\cn/G$ to be the moduli space of flat
connections.  Let $m^I,~ I=1,\cdots, \dim{\cm}$ be local coordinates
on $\cm$.  Flat connections are then parameterized as $A(m)$.  Given
two nearby flat connections as $A(m)$ and $A(m+d m)$, we expand the
latter to see that the condition for it to also be a flat connection
is that
\be d_A\frac{\partial A}{\partial m^I}dm^I\seq0\ \ .\ee
By
definition, the zero-mode of the $(1,1)$ curvature component on
$P\times\ca$, $\psi^{(0)}$, satisfies the equation
\be
d_A\psi^{(0)}\seq 0\ ,\ee
where $A$ is a flat connection.  Thus we
immediately find a basis from which $\psi^{(0)}=\psi^{(0)}_I dm^I$
may be constructed; namely, $\psi^{(0)}_I=\frac{\partial A}{\partial
m^I}$.

We seek observables in the \bft{} which we can formally write
in terms of $\frac{\partial A}{\partial m^I}$ assuming we have chosen
a coordinate patch on $\cn$.  In order for them to be observables
they must be gauge invariant and diffeomorphism invariant.  These
conditions are related as we will soon see.  Let us now turn to their
construction.

For a homology two-cycle, $\Gamma$, on $M$, we define
\be
\co_{IJ}~\equiv~\half \int_{\Gamma} Tr ( \frac {\partial A}
{\partial m^I}\wedge
 \frac {\partial A}{\partial m^J} ) \ \ .\ee
Under a gauge transformation, $A\to A^g$, $\frac{\partial A}
{\partial m^I}$ transforms into $\frac{\partial A^g}{\partial m^I}$.
Then for an infinitesimal gauge transformation, with parameter $ \epsilon $,
\be
 \delta_{\epsilon} \co_{IJ}\seq  \int_{\Gamma}
 Tr(\frac {\partial \epsilon}{\partial m^{[I}}
 \frac {\partial F}{\partial m^{J]}}) \ \ .\ee
Thus we see that $\co_{IJ}$ is gauge invariant if $A$ is a
flat connection.  Hence it is a possible observable in  \bfts.

A check of diffeomorphism invariance remains to be done.
Diffeomorphisms of the manifold, $M$,  by the vector field, $K$,
are generated by the Lie derivative $\cl_K=d i_K + i_K d$.
By direct computation,
\be
\cl_K\frac {\partial A}{\partial m^I}\seq i_K d_A\frac {\partial A}
{\partial m^I} \pls [\frac {\partial A}{\partial m^I},\alpha(K)]
\pls d_A(i_K\frac {\partial A}{\partial m^I})\ ,\ee
where $\alpha(K)\equiv i_KA $.  If $A$ is a flat connection,
the first term in the right-hand-side of this expression vanishes.
The second term is a gauge transformation.  Although the last term is
inhomogeneous the fact that it is a total derivative means that after
integrating by parts and imposing the flat connection condition, its
contribution vanishes.
It then follows that  $\co_{IJ}$  is an example of a gauge invariant
operator whose correlation functions in the \bft{} are diffeomorphism
invariants.

Having convinced ourselves, by the example above, of the existence of
operators in \bfts{} which lead to diffeomorphism invariant
correlation functions, we must now establish a  procedure for
constructing such quantities.   This will be done by implementing
the ideas in section \ref{Sec_Idea}; namely,  almost compute the
topological  invariants from the \sbft{} theory.  By almost, we
mean integrating over all of the fields in the functional integrals
except for the gauge connection and the field $B$.  This will leave
us with a functional integral expression over the space of fields in
the ordinary $BF$ theory but with operator insertions at various
points on the manifold.  Now we know that we are in fact computing
topological invariants.  Then it follows that these operators, which
will appear as functionals of the connection will be
physical correlation densities\footnote{The relation of these
quantities to what we normally
expect observables to be is discussed in section \ref{Sec_Idea}.} in the
\bft{} whose expectation values are
topological invariants.

To illustrate the procedure, let  us write the generic Donaldson
polynomials as $\co_i(\phi,\psi,F;\cc_i))$ where $\cc_i$ is the cycle the
observable is integrated over.  Then we
have to compute
\bea
\ev{\prod_i \co_i(\phi,\psi,F;\cc_i)}_{SBF}
\seq&& (Z_{SBF})^{-1}
\int [d\mu]_{SBF}~e^{-S_{SBF}}\times\nen
&&\times~\prod_i \co_i(\phi,\psi,F;\cc_i)  \ \ ,\label{CORFCN_PSI}
\eea
where $Z_{SBF}$ is the partition function of the super-$BF$ theory
and $[d\mu]_{SBF}$ (see appendices \ref{A_SBF} and \ref{A_Nota})
is the measure for the
path integral over the fields
$\chi,\psi$, etc. with the $\chi$ zero-modes inserted.
It is known (see  appendix \ref{A_SBF}) that certain classes of operators
$\co_i$ exist for which these correlation functions are topological
invariants.

The integral over
$\lambda$ may be performed leading to the delta function $\delta
(\Delta_A\phi +[\psi,\dual\psi])$.  This means that the $\phi^a$
field is replaced by $\ev{\phi^a(x)}=-\int_MG^{ab}(x,y)
[\psi,\dual\psi]_b$, where $G^{ab}(x,y)$ is the Green's function
of the scalar covariant laplacian ($\Delta_A$), in the computation
of the correlation function of the observables.  Since the
functional integral has support only on flat connections,
if there are no $B$-fields in the observables (as is the case
for our $\co$'s), the correlation densities
reduce effectively to functions of $\psi$ and the flat connections only.
In order for  correlation functions to be non-zero, the product
of the observables -- reduced in this way -- must include all
$\psi$ zero-modes.
For a genus $g\geq2$ handlebody,  this number is $(g-1)\dim(G)$.
Let us now look at the various classes of correlation functions.

The vacuum expectation value of a single $\co$ is a topological
invariant in the \sbft.{}  Hence, the gauge invariant
operator ($S_S$ is defined in appendix \ref{A_SBF})
\be
\hat\co(A;\cc)~\equiv~ \int [d \mu]_S e^{-S_S}
\co(\phi,\psi,F;\cc) \ \ ,\ee
has the property that its vacuum expectation value in the \bft{} is a
topological invariant.  It is important to note that in general,
$\hat \co$ depends on the
background metric on $M$.  Furthermore, although it is gauge
invariant, it is not in the cohomology of the, $Q_{SBF}=Q^H+Q_{YM}$, total
BRST charge  (see  appendix \ref{A_SBF} for a discussion on
$Q^{H}$), where $Q_{YM}$ is the Yang-Mills BRST charge.  Hence, its
correlation functions will not be
independent of the background metric, in general.  Additionally,
the $\hat\co$ will be non-local operators in general.  Although
these last two points may be viewed as drawbacks of this approach,
there is one important lesson to be learned here.  This
construction clearly demonstrated that the three dimensional analogs of
Donaldson invariants give
rise to operators in the \bfts{} whose vacuum expectation values,
in the latter theories, are themselves topological and are physical
in three dimensional quantum gravity.
It should also be noted that although the fields $B$, $c$,
$\bar c$, $c'$, ${\bar c}'$ appear in $S_S$,
they do not survive the $[d\mu]_S$
integration due to $\psi$ zero-mode saturation.

Until this point, we have only looked at the vacuum expectation
values of the $\hat\co$'s.  Now, we would like to investigate the
expectation value of $\hat \co$ in any physical state of the \bft.
In particular, we would like to see whether or not such an expression
is independent of the background metric, $g_{\mu\nu}$, used in
forming the gauge fixed action.  Let us suppose that such a state may
be constructed out of the action of Wilson loop operators on the
vacuum.  Alternatively, we can ask whether or not the correlation
function of the $\hat \co$'s with Wilson loop operators,
$W[R,\gamma]= Tr_R P\exp{(\oint_\gamma A)}$, is background metric
dependent.  Hence we are led to study the functional integral

\bea
 {\cal E}(R,\gamma,\cc) &\seq& \int [d\mu]_{SBF} e^{ -S_{SBF}}
\co(\phi,\psi,F;\cc)  W[R,\gamma]\nen
&\seq& \int [d\mu]_{BF} e^{-S_{BF}} \hat\co(A;\cc)
W[R,\gamma]\ \  .\eea
Functionally differentiating ${\cal E}(R,\gamma,\cc)$ with respect
to the inverse metric, $g^{\mu\nu}$,  we find
\be
\frac{\delta {\cal E}(R,\gamma,\cc)}{\delta g^{\mu\nu}} \seq
\int [d\mu]_{SBF} e^{ -S_{SBF}}\Lambda_{\mu\nu} \co(\phi,\psi,F;\cc)
Tr_R P(\oint_\gamma\psi e^{\oint_\gamma A})\ \ ,\ee
after use of the properties of $S_{SBF}$ and where $\frac{\delta
S_{SBF}}{\delta g^{\mu\nu}}=\{Q,\Lambda_{\mu\nu}\}$ with
\be
\Lambda_{\mu\nu} \seq
\frac{\delta }{\delta g^{\mu\nu}} \int_M (\lambda d_{A}\dual \psi +
\lambda ' d_{A}\dual \chi +{\bar c}'d_{A}\dual B + \tilde{c}
\delta A ) \ \ .\ee
We  notice that the integral over $ \phi ' $ yields
$ \delta ( \Delta_{A}{}^{(0)} \lambda ') $.  Since we assume that
$ \Delta_{A}{}^{(0)} $ does not
have any zero-modes then this restricts $ \lambda '$ to be zero.
As a result of this, the only appearance of $\chi$ left is in the
action.  Integrating over this field we find $\delta(d_A\psi-\dual
d_A\eta)$.  Now, the integrability condition for this restriction
is $[F,\psi]=\nabla_A{}^{(0)}\eta$.  However, as the integral over
$B$ can be seen to enforce $F=0$, we find that $\eta=0$, hence
$d_A\psi=0$.  This means that all $\psi$'s in the path integral
are now restricted to be zero-modes.  For all but the first term in
$\Lambda_{\mu\nu}$, the $\lambda$ integration can be performed and it
restricts each $\phi$ in the $\co$
to be replaced by an expression (see below) which
depends on two $\psi$ zero-modes.  This then means that the path
integrals involving each of the last three terms in $\Lambda_{\mu\nu}$
is saturated by $\psi$ zero-modes due to the presence of
$\co$.  Thus, we see that the extra $\oint_\gamma \psi$ due to
the Wilson loop makes those expressions vanish.  We are then left
with the first contribution for $\Lambda_{\mu\nu}$.  If $\co$ depends
on $\phi$ this will not be zero.  Thus we deduce that $\frac{\delta
{\cal E}(R,\gamma,\cc)}{\delta g^{\mu\nu}}=0$, in general, only if
the $\co$ does not depend on $\phi$; otherwise, the
only restriction on $\co$ is that it saturates the
number of $\psi$ zero-modes.   Additionally, the result will not
be altered if we included more than one Wilson loop in $\cal E$.
Thus we conclude that the correlation functions of those $\hat
\co$ operators whose ancestors -- $\co$ -- saturated the number
of fermion zero-modes and are independent of $\phi$,
with Wilson loops is independent of the
background metric.

Observables in the
\bft{} which depend on $B$ have been constructed in the literature
\cite{Broda,TFT_rev}.   An immediate observation is that if we compute
correlation function
of quantities which depend on $B$ then the path integral is not
restricted to
$ \cal N $. This invalidates the proof above. However, if we
restrict to
$ M = \Sigma \times R $ ( i.e canonical quantization ), then there
will be
only dependence on $ B|_\Sigma $ in the observables and the
restriction
to $ F|_\Sigma = 0 $ survives. In this case correlation functions
involving $A$, $B$ and
$\hat\co(\psi)$ are gauge invariant and metric independent.

Haven given formal expressions for physical correlation functions in
\bfts, we would now like to trace our steps back to the
analysis at the beginning of this section and see how it might
arise directly from \sbfts. Let us choose quantum gravity on a
genus three handle body as a specific theory; thus,
$g=3$ and $G=SO(2,1)$.  Six $\psi$ zero modes are needed so we
pick three homology $2$-cycles which we label as $\Gamma_i$.  Then we
compute the correlation function $\ev{\prod_{i=1}^3\int_{\Gamma_i}
Tr(\psi\wedge\psi)}_{SBF}$, in the \sbft{} also with $g=3$ and $G=SO(2,1)$.
After integrating out the $Y$-fields, we obtain\footnote{In general, the path
integrations over the bosonic zero-modes are understood to drop out due to the
division by $Z_{SBF}$ in the expressions for the correlation densities.}
\be
\ev{\prod_{i=1}^3\int_{\Gamma_i} Tr(\psi\wedge\psi)}_{SBF}\seq
(Z_{SBF})^{-1} \int [d\mu]_{BF,\alpha_1\ldots\alpha_6}~
e^{-(S_{BF}+S_{BF,gf})}
{\hat\co}(A)\ \ ,\ee
\pagebreak
where
\bea
{\hat\co}^{\alpha_1\cdots\alpha_6}(A)\seq&&
T(A)\int_{\Gamma_1} Tr(\Upsilon^{[\alpha_1}(A)\wedge\Upsilon^{\alpha_2}(A) )
{}~\times\nen &&\times~\int_{\Gamma_2}
Tr(\Upsilon^{\alpha_3}(A)\wedge\Upsilon^{\alpha_4}(A) )~\times\nen
&&\times~\int_{\Gamma_3}
Tr(\Upsilon^{\alpha_5}(A)\wedge\Upsilon^{\alpha_6]}(A) )\ \
.\label{PsiPsi}\eea
Here, the $\Upsilon^{\alpha_i}(A)$ form a six-dimensional basis for
$H^1(M,G)$.  By $[d\mu]_{BF,\alpha_1\ldots\alpha_6}$ we mean
$[d\mu]_{BF}$ with the functional measure over flat connections,
$A^{(0)}$ given by $[d A^{(0)}_{\alpha_1}]\cdots [d
A^{(0)}_{\alpha_6}]$. The $A^{(0)}_{\alpha_i}$ and
$\Upsilon^{\alpha_i}(A)$ are chosen to form a canonical basis for
$T^*\cm$ as in
ref. \cite{Wit(TYM)}.
As this expression was derived directly
from the super-$BF$ theory, the result is independent of the choice
of basis for the fermionic zero-modes.  The quantity $T(A)$ arises
from the non-zero mode integration in $[d\mu]_S$.
The remaining functional
integral has support only on flat connections, hence $T(A)$ is
ostensibly the Ray-Singer (R-S) torsion \cite{RaySing}.  We then
identify the $\Upsilon(A)$ as $\frac{\partial A}{\partial m}$.
Notice that in our analysis of the \bft{} at the beginning of this
section, it was not evident that the R-S torsion appears as part of
the observable's definition.

Now we realize that
\be
\ev{\prod_{i=1}^3 \int_{\Gamma_i}
Tr(\psi\wedge\psi)}_{SBF}\seq \ev{{\hat\co}(A)}_{BF}\ \ .\ee
Then interpreting ${\hat\co}(A)$ as a correlation density in the
\bft{} we continue the computation to find
\be
\ev{{\hat\co}(A)}\seq(Z_{SBF})^{-1} \int_\cn ~ \prod_{i=1}^3
\int_{\Gamma_i} Tr(\Upsilon(A^{(0)})\wedge\Upsilon(A^{(0)}))\ \ ,\ee
where $\Upsilon(A^{(0)})$ is a form on the space, $\cn$, of connections.
The functional integral, $\int_\cn$ over $\cn$ is done with a wedge
product of the $\Upsilon$'s, on that space, understood.

As a second example, we construct a correlation density in quantum gravity
which is considerably less obvious in the $BF$ theory than the prior
example.  We start with $\int_\gamma Tr(\phi\psi)$, here
$\gamma$ is a one-cycle.  It carries ghost number three.  Thus we
construct a correlation density in quantum gravity on a genus $g\geq2$
handle-body given as
\be
\hat\co(A;\gamma_i)\seq (Z_{SBF})^{-1} \int [d\mu]_S
e^{-S_{S}}
\prod_i\int_{\gamma_i} Tr(\phi\psi)\ \ .\ee
Integrating over $\lambda$  we find that at the expense of a factor
$\det{}^{-1}(\triangle_A^{(0)})$, we should replace $\phi(x)$ by
$-\int_{M_y} G_A(x,y)[\psi(y),\dual\psi(y)]$, where $G_A$ is the
Greens' function of the scalar covariant laplacian.  Then
functionally integrating over $\psi$ we obtain
\bea
\hat\co(A,\gamma_i)&\seq& -(Z_{SBF})^{-1} T(A)~\times \nen
&~\times&\prod_i\{\int_{\gamma_i} Tr\{\int_{M_y} G_A(x_i,y)[\Upsilon(A(y)),
\dual\Upsilon(A(y))]\Upsilon(A(x_i))\}\}\nen
&&\eea
to be another correlation density in quantum gravity.  In this expression,
the $\Upsilon$'s appear anti-symmetrized as in (\ref{PsiPsi}).

Concluding this sub-section, we note one more point about
the correlation densities we have been writing down.  Unlike
observables, our expressions are, in addition to
being non-local in the \bft,{} given in terms of path integrals.
These functional integrals are best computed in perturbation theory.
However, by invoking BRST theorems we were able to obtain some
expressions non-perturbatively, in the above.  It is safe to say
that one lesson we have learned from this sub-section is that for
diffeomorphism invariant theories, quantum gravity in particular,
we must enlarge our scope of what an observable is.  Here, we have
used the geometry of the universal bundle and more directly the
de Rham complex on moduli space to guide us.  Presumably, this
direction is worth a try in four dimensions also.  We will turn
to the latter in the next section.  However, before that, we
would like to discuss some  even more profitable  results; namely,
expressions for wavefunctionals based on the universal bundle
geometry.

\subsection{Wavefunctionals}\label{SS_BFWfcn}
\setcounter{equation}{0}
\medskip
\forcepar
The physical Hilbert space of a \sbft{} consists of
$L^2$-functions on the moduli space, $\cal M$, of flat connections.
In principle, quantization of this field theory is then reduced to
quantum mechanics on $\cal M$.  However, the pragmatism of such a
program is limited as, a priori,  it becomes unwieldy to pull such
wavefunctions back into wavefunctional of the connection.  In this
sub-section we will demonstrate how this problem may be obviated.
To be precise, we will write down expressions for the functionals of
the connections on the $G$-bundle which are annihilated by the constraints
of the theory.

Correlation functions of observables in TQFT's are equal to the integral
over moduli space of a top form on that space \cite{Wit(TYM)}.
Typically, such a
top form is wedge product of forms of lesser degree:
\be
\ev{\prod_i^d {\cal O}_1}\seq \int_\cm \Psi\ \ ,\qquad \Psi\seq
f_1\wedge f_2\wedge\cdots\wedge f_d\ \ ,\ee
where the forms, $f_i$, are obtained after integrating over the non-zero
modes and fermionic zero-modes in the path integral.  Now, let us assume
that a metric exists on $\cm$ so that we can define the Hodge dual map
which we  denote by the tilde symbol.  Then  $\tilde \Psi$ is  a
scalar function on moduli space.  Let us now give representations for
$\Psi$.  All we seek is $\Psi$'s which are gauge invariant and have
support only on flat  connections, $\omega$,  on $\Sigma_g$: $\Psi[\omega]$.

Clearly \cite{Wit(TopChnge)}, a delta function, $\delta(F)$,
where $F$ is the curvature of a
$G$-bundle over $\Sigma_g$ satisfies our criteria.  However, as it is
highly unlikely to be normalizable.  Regardless, we
realize that it might be worthwhile to look at diffeomorphism invariant
theories on $\Sigma_g$ which are defined in a flat connection background.
Considering the two-dimensional \bft{} we see that the analog
of $B$, $\varphi$, is an $ad(G)$-valued zero form and the action is
$S_{BF}^{2D}=\int_{\Sigma_g} Tr(\varphi F)$.
The delta function arises from the integral over $\varphi$.
If we constructed the analog of $S_{SBF}$, we would find that
it shares many of the terms which appear in the three
dimensional action. However, here the analog of $\chi$ (we will
call this $\xi$ below) is
a zero-form.  What is more,  there are no primed fields due to the
degree of $\xi$.  Considering this, we introduce
the functional $\int_{\Sigma_g}Tr( \xi d_{\omega} \psi)$, for
$ad(G)$-valued, Grassmann-odd
zero- ($\xi$) and  one- ($\psi$) forms defined in a flat connection
background, $\omega$, It is invariant under the local symmetry,
$\delta\psi=\epsilon d_\omega\phi$ and upon gauge fixing it we obtain
the quantum functional
\be
S_\omega\seq \int_{\Sigma_g}Tr(\xi d_\omega\psi\mi \eta
d_\omega\dual\psi\mi
\lambda\dual\triangle_\omega\phi)\ \ .\ee
The partition function for this action is metric independent as the
part of $S_\omega$ which is metric dependent
is exact with respect
to the BRST charge for the gauge fixing of the symmetry just discussed.
Furthermore, it is simple enough to compute exactly and is found to be
equal to the Ray-Singer torsion of the $G$-bundle with flat connection,
$\omega$.  In fact, the action $S_\omega$ is recognized as the action for
a two-dimensional Grassmann-odd BF field theory in a flat connection
background.
As was the case with the \sbft{}, the correlation functions of
quantities such as $\half Tr(\phi^2(x))$, etc., are topological
invariants.   This is seen to be due to the transformation given by the
BRST charge: $\{Q,\psi\}=
d_\omega\phi$.  The partition function has support only on solutions of
those $\psi$ which are in $\ker(d_\omega)$.  Hence, they span the cotangent
space of $\cm(\Sigma_g)$ whose dimension is $(2g-2)\dim(G)$.

As before, let us focus on three dimensional quantum gravity taking
$G=SO(2,1)$. Our
first example of a wavefunctional is found by taking the $(3g-3)$
times product of
$\int_{\Gamma_i}
Tr( \psi\wedge\psi)$ where the $\Gamma_i$ are homology 2-cycles in
$\Sigma_g$:
\be
\Psi[\omega]\seq \int [d\xi][d\psi][d\eta][d\lambda][d\phi] e^{-S_{\omega}}
\prod_{i=1}{}^{(3g-3)}\int_{\Gamma_i}Tr( \psi\wedge\psi)\ \ ,\ee
defined over the two-dimensional super-$BF$ theory.  The generic form of
the wavefunctionals obtained by this construction is
\bea
\Psi_{\vec n}[\omega] &\seq& \int[d\xi][d\psi][d\eta][d\lambda][d\phi]
e^{-S_{\omega}} \prod_{i=1}^{n_4}
Tr(\phi^2(x_i)) ~\times\nen
&&~\times \prod_{j=1}^{n_3} \oint_{\gamma_j}Tr(\phi\psi)
\prod_{k=1}^{n_2}
\int_{\Gamma_k}Tr( \psi\wedge\psi)\ ,\eea
subject to the condition $4n_4+3n_3+2n_2=\dim{\cm(\Sigma_g,G)}$.  If $\omega$
is not an irreducible connection, then there are no $\phi$ zero-modes
and the only non-zero $\Psi_{\vec n}[\omega]$ are those for which
$n_4=n_3=0$.

In the preceding, we have not used the full power of the two-dimensional
\sbft{}.  As a matter of fact, we did not use it at all.  The transition
from $S_\omega$ to the \sbft{} on $\Sigma_g$ is straightforward.  Its action
is
\be
S_{SBF}^{2D}\seq \int_{\Sigma_g}Tr(\varphi F)\mi S_\omega \pls
\int_{\Sigma_g} Tr(\lambda[\psi,\dual\psi])\ \ ,\ee
where $\varphi$ is a zero-form which imposes the flat connection condition
on $\omega$ and the rest of the action is reminiscent of the three dimensional
theory but without the primed fields.  Unlike the pure $S_\omega$ theory,
the absence of $\phi$ zero-modes does not imply $\phi=0$, but
$\phi(x)=-\int_{\Sigma_{g,y}} G_A(x,y)[\psi(y),\dual\psi(y)]$ as we saw
in the previous sub-section.  Thus, more wavefunctionals result from this
theory.  They are of the same form as $\Psi_{\vec n}$ except that the
functional measure must be enlarged to include all the fields in
$S_{SBF}^{2D}$.  Additionally,
$S_\omega$ is replaced by $S_{SBF}^{2D}$.
We find the general form of these wavefunctionals
to be
\bea
\Psi_{\vec n}^S[\omega] &\seq&  (-)^{n_3}\prod_i^{n_4}
Tr((\int_{\Sigma_{g,y}} G_\omega(x_i,y)[q(y),\dual q(y)])^2)~\times\nen
&~\times&\prod_j^{n_3} \oint_{\gamma_j}Tr(\int_{\Sigma_{g,y}} G_\omega(x_j,y)
[q(y),\dual q(y)] q(x_j))~\times\nen
&~\times& \prod_k^{n_2}
\int_{\Gamma_k}Tr(  q\wedge q)\ ,\label{CanWfcn}\eea
again with $n_4+n_3+n_2=\dim(\cm(\Sigma_g,G))$ and where the
$q(\omega)$ form a $(2g-2)\dim(G)$ dimensional basis for
$H^1(\Sigma_g,G)$.  As was the case in eqn. (\ref{PsiPsi}), the $q$'s
appear in a totally anti-symmetric combination.

Our philosophy thus far has been to identify a Riemann surface
(which is homeomorphic
to the hypersurface of the foliated three-dimensional
\bft{}) and construct a servant partition
function\footnote{We will call these servant partition functions
to distinguish them from the partition functions of the theories we
are constructing the wavefunctionals for.}
for fields in a background which solves the constraints of the
three-dimensional \bft{}.  Having done this we then identified
operators which yield diffeomorphism invariant observables in the
two-dimensional topological ``theory".  We assume that we can solve
the equation which defines the constraints (as though they were classical
equations) and parametrize them by the coordinates on moduli space.
For example, the $\omega$ which defines the background above is really
$\omega(x;m)$.  This means that the wavefunctionals are not simply defined
at one point in moduli space, but rather on all of $\cm(\Sigma_g,G)$.
We advocate this as a very robust approach to constructing quantum gravity
wavefunctionals as (1) we need only solve the constraints classically and (2),
thanks to our experience with TFT's it is rather straightforward to
at least formally construct the servant partition functions and
correlation functions in such a parametrized background.

Now, the fact that, in the previous sub-section, we were successful in
formally constructing correlation densities leads to another possible
approach to
constructing wavefunctionals.  If those correlation densities can be written,
as operators, as $\hat \co={\hat O}^\dagger{\hat O}$ for some operator
${\hat O}$ and adjoint, $\dagger$, then we would have $\ev{\hat\co}=
\langle0|{\hat O}^\dagger{\hat O}|0\rangle$.  Interpreting this as the
norm of a state ${\hat O}|0\rangle$, it is suggestive that the
wavefunctional of such a state may be formed from the path integral
expression for the correlation density.  The manner in which we see this state
arising is analogous to sewing in the \sbft{}.  Hence, we expect to be
able to form the corresponding wavefunctional by surgery in the \sbft{}.
Although we delay detailed investigation of such an approach until
a future publication, we would like to point out here that normalized
wavefunctionals are expected.  In the rest of this sub-section,
we will show how to construct
wavefunctionals from a \sbft{} on a three
manifold whose boundary is $\Sigma_g$.

We start with a  \sbft{} for a $G$-bundle over a three-dimensional manifold
$M$ with boundary $\Sigma_g$.  Then we insert the pertinent operators as
was done in the previous sub-section.  Having done this, we choose a
polarization (for which the fields in the geometrical sector appear as
``position" variables) and perform all functional
integrals with appropriate boundary conditions.  This gives a wavefunctional
for the \sbft{} which is annihilated by the full BRST operator.   It is also
gauge and diffeomorphism invariant \cite{Wit(TYM),Thom}.  In general, there may
be fields which do not depend on the boundary values of the geometrical sector.
 Starting with the wavefunctional of the \sbft{}, we integrate over their
boundary values.
This leads to a functional, $\Psi[\omega,\varpi]$, where $\omega$ is a
flat connection on $\Sigma_g$, $\varpi$ denotes  a zero-mode of $\psi$ and is a
solution of
$d_\omega\varpi=0$ on $\Sigma_g$, and we have replaced the boundary
value of $\phi$ with the appropriate expression in terms of $\omega$ and
$\varpi$.  $\Psi$ also depends on the boundary values of $c$ and
$c'$; however, for notational simplicity they will be omitted.

Focusing our attention on genus-g handle-bodies, $\partial M=\Sigma_g$,
we compute the correlation functions
for the topological invariants fixing the boundary value of the connection
to be  a flat connection on $\Sigma_g$.  This can be done by inserting a
delta function, $\delta(U_I(A)-
g_I(\omega))$ for each longitude, $l_I$ in $M$.  Here, $U_I(A)$ is the
holonomy of the connection along the longitude, $l_I$, and $g_I(\omega)$
is the holonomy of a parametrized flat connection on the cycle,
$b_I$, on $\Sigma_g$
which (in the handlebody) is homotopic to $l_I$.
For the $\psi$ field, we insert delta functions $\delta(\oint_{l_I}\psi
-\oint_{b_I}\varpi)$ where $l_I$ and $b_I$ are as before.
Consequently, we arrive at our  generic ansatz for wavefunctionals of
\sbfts{}:
\bea
\Psi[\omega,\varpi]&=&\int
[d\mu]_{SBF}\! \prod_{I=1}^{(g-1)\dim(G)}\hskip -20pt\delta(U_I(A)-
g_I(\omega))\! \prod_{J=1}^{(g-1)\dim(G)}\hskip -20pt\delta(\oint_{l_J}\psi
-\oint_{b_J}\varpi)\times\nen
&~&\hskip 45pt\times ~\prod_i\co_i(\phi,\psi,F;\cc_i)~e^{-S_{SBF}}\ \ .
\eea
In this expression, the product of polynomials, $\prod_i\co_i$ is such that
it saturates the number of $\psi$ zero-modes.

Now we must project
out $\varpi$.  This we do by  treating $\Psi[\omega,\varpi]$ as a
superfield and obtaining the wavefunctional of $\omega$ as a
component via
superfield projection.  Choose a
$(2g-2)\dim(G)$-dimensional basis, $q_\alpha(\omega)$,
for $H^1(\Sigma_g,G)$ and expand the one-form field, $\varpi$, in it as
$\varpi\equiv \sum_\alpha \theta^\alpha q_\alpha$, where the Grassmann-odd
coefficients $\theta^\alpha$ are the quantum mechanical oscillators.
As the wavefunctional  depends on $H^1(M,G)$ only $(g-1)\dim(G)$ of the
$\theta^\alpha$ will be non-zero.
Then, $\Psi$ is formally $\Psi[\omega,\theta]$ and we write,
\be
\Psi_{\alpha_1,\ldots,\alpha_n}[\omega]\seq
\frac{\partial^n}{\partial \theta^{\alpha_n}\cdots\partial
\theta^{\alpha_1}}\Psi[\omega,\theta]{\mid}\ \ ,\ee
where the slash means setting $\theta=0$ after differentiating.
Each $\Psi_{\alpha_1,\ldots,\alpha_n}$ for $n=1,\ldots,(g-1)\dim(G)$,
is a  wavefunctional in the \bft{} in that it satisfies the
constraints of the latter theory.   We adopt this procedure as it is closest
to the sewing/surgery procedure, is of geometrical significance (see
below)  and it incorporates the two naive guesses:
setting $\varpi=0$ or  integrating out $\varpi$.

The closest analogy of these $\Psi_{\alpha_1,\ldots,\alpha_n}[\omega]$ is to
Hartle-Hawking wavefunctionals \cite{HarHawk}.  Notice that unlike the
previous wavefunctionals, $\Psi^S_{\vec n}$, in eqn. (\ref{CanWfcn}),
which are analogous to canonical wavefunctionals, these are
functionals only of half of the flat connections on $\Sigma_g$.  This
is due to the fact that the meridians of the handlebody are
contractible in $M$.

Based on our discussion at the beginning of this sub-section, we
realize that the $\Psi_{\alpha_1,\ldots,\alpha_n}$ are $n$-forms on
moduli space.
This returns us to our earlier discussion of the wavefunctionals of
quantum gravity as being $L^2$-functions on moduli space.  In writing
down the $\Psi_{\alpha_1,\ldots,\alpha_n}$, we have given formal
expressions for the pertinent functions on the moduli space in terms
of the physical variable in the problem; namely, the connection.

The question remains which of them is normalizable.  Although we do
not have much to say about this question in this work, we would like
to bring to the fore a possible strategy for normalizing
wavefunctionals constructed in this way.  Any closed piecewise linear
three dimensional manifold, $N$, may we formed via the Heegaard
splitting, $N=M_1\cup_h M_2$, where $M_1$ and $M_2$ are two
handlebodies whose boundaries are homeomorphic (with map, $h$) to
each other \cite{Rolf}.  Making use of this, we view \cite{BroLif} the norm of
$\Psi[\omega,\theta]$ to be a functional integral on $N$,
$\Psi[\omega,\theta]$ itself to be the functional integral on $M_1$
and its adjoint to be the functional integral on $M_2$.  Reversing,
we start with the functional integral for the correlation functions
of polynomial invariants on $N$, perform surgery and then identify
that component of $\Psi$ which appears in the form
$\Psi[\omega]^\dagger\Psi[\omega]$ after integrating $[d\mu]_S$.
Obtaining the adjoint ($^\dagger$) is interpreted as arising from the
surgery/sewing process.

\vskip 0.5truein
\section{Application to 4D Quantum Gravity}\label{Sec_4D}
\setcounter{equation}{0}
\medskip
\forcepar
In this section, we sketch how the above construction can be
 realized in four dimension quantum gravity. The main ingredient that we
have to supply is a TQFT whose action starts off as an Einstein-Hilbert
action, or rather when some of the fields are put to zero one gets the usual
action for gravity. Generally, we expect that there is more than one such
 action.  Unlike the three dimensional case where quantum gravity
was already defined
over a finite dimensional phase space (flat connections modulo gauge
 transformation), in four dimension there are propagating fields and we can
then try projecting the theory onto many different moduli spaces.
In a way, the Einstein-Hilbert action is a good example for this construction,
Whereas it is non-renormalizable, the topological
projection takes us to a renormalizable theory in which to do
calculations on physical observables for quantum gravity.
In addition,
we can construct formal, diffeomorphism invariant
expressions\footnote{We remind the reader that in constructing
observables, we work in covariant -- not canonical -- quantization.}
without integration over the whole manifold.

Our first task is to select a TQFT. The logical candidates are
topological gravity theories.  Four dimensional topological gravity
theories for which the pure metric part of the action is given by the square
of the Weyl tensor, not the Einstein-Hilbert action, are known
\cite{(WeylTG)}.  As mentioned above, we seek a topological gravity
theory whose pure metric action is the Einstein-Hilbert action.  Now,
TQFT's may be obtained from supersymmetric theories via twisting
\cite{Wit(TYM)}.  The fact that the four dimensional gravity theories
which were first constructed were conformal is apparently correlated
with the fact that N=2 supergravity in four dimensions has this feature.
A Poincar{\' e} supergravity theory was proposed sometime ago by de
Wit\footnote{We thank S. J. Gates, Jr. for bring this to our
attention.} \cite{deWit}.  Thus we expect that a twisted version of
this should exist as a topological gravity theory.

In ref. \cite{AnsFre}, a topological gravity theory with
Einstein-Hilbert action as is pure metric part was obtained by
twisting a N=2 supergravity theory. Here, we will simply use
the results of this work.
The twisting procedure defined a Lorentz scalar, Grassmann-odd (BRST)
charge, $Q$ which is nilpotent.
As it turns out this topological gravity theory is seen to be the
TQFT for the projection of the spin-connection form (in the second order
formulation) to be self-dual:
\be
w^{-ab}(e)\seq 0\ \ ,\ee
where $a$ etc. are Lorentz indices.

The observables are constructed from the cohomology of Q. After some
 re-definitions of the fields one ends up with a BRST charge whose action
 upon the geometrical sector of the theory is
\bea
&&Q: e^{a} \to\psi^{a} \mi \cd \epsilon^{a} \pls \epsilon^{ab} \wedge
e_{b}\ \ ,\nen
&&Q:w^{ab} \to \chi^{ab}\mi \cd \epsilon^{ab}\ \ ,\nen
&&Q: \psi^{a} \to-\cd \phi^{a} \mi \eta^{ab} \wedge e_{b} \mi\chi^{ab} \wedge
\epsilon_{b} \pls\epsilon^{ab} \wedge \psi_{b}\ \ ,\nen
&&Q: \phi^{a} \to \epsilon^{ab} \wedge \phi_{b}\to\eta^{ab} \wedge
\epsilon_{b} \ \ ,\nen
&&Q: \chi^{ab} \to-\cd \eta^{ab} \pls\epsilon^{ac} \wedge \chi_{c}^{b}
\mi\chi^{ac} \wedge \epsilon_{c}^{b}\ \ ,\nen
&&Q: \eta^{ab} \to \epsilon^{ac} \wedge \eta_{c}^{b}
\mi \eta^{ac} \wedge \epsilon_{c}^{b}\ \ ,\nen
&&Q: \epsilon^{a}\to \phi^{a}\pls\epsilon^{ab} \wedge \epsilon_{b}\ \
,\nen
&&Q: \epsilon^{ab}\to \eta^{ab}\pls \epsilon^{a}_{c} \wedge
\epsilon^{cb}\ \ ,\eea
where $ \epsilon^{ab}$ and $ \epsilon^{a} $ are the ghosts for Lorentz and
diffeomorphism symmetries, respectively.
 As was mentioned above all this is in second order
formalism. Although in the BRST transformations all the fields look
 independent,
this is not the case. However, according to \cite{AnsFre},
these transformations are consistent with
the conditions of the second order formalism :
$ w^{ab} \wedge e_{b} = de^{a}$ ,
 $\chi^{ab} \wedge e_{b}=-\cd \psi^{a}-R^{ab} \wedge \epsilon_{b} $ , etc...

The cohomology is constructed exactly as in Donaldson-Witten theory
\cite{TFT_rev}
and observables are found.  For example,
\bea
\co^{(4)}&\seq& \half Tr (\eta^2)\ \ ,\nen
\co^{(3)}&\seq&\int_{\g}Tr (\eta  \chi )\ \ ,\nen
\co^{(2)}&\seq&\int_{\Gamma} Tr (\eta R \pls \half\chi \wedge \chi )\
\ ,\nen
\co^{(0)}&\seq&\int_{M} Tr(R \wedge R )\ \ .\eea
The geometrical meaning of each field is in pure analogy with those in
 Donaldson-Witten theory.
As in the three dimensional case, the correlation functions of these
observables become,
after integration over the non-zero-mode parts of the fields, functions of the
zero modes of $ \chi^{ab}$ and the veirbein, $e^a$.  The zero-modes
of $\chi$ are a basis for the cotangent space to
the moduli space of $ w^{-ab}(e)=0$.

Now that we have described the results of \cite{AnsFre}
 we would like to indicate
how the constructions of the previous sections can be applied here.
To construct the observables we insert, in the path integral a combination
of the TQFT observables which saturates the $\chi$ zero-modes and then
integrate over all the fields except
$e ,\epsilon^{a} $ and $\epsilon^{ab}$. This will result in a non-local
expression in terms of the veirbein whose expectation value
is a topological invariant of space time.  In this way, we obtain
$\hat\co(e)$ from the $\co^{(k)}$ above.

The wavefunctional construction is also very similar. However, at
present, we can
only construct the Hartle-Hawking type wavefunctionals, as unlike the
3D case, we do not have
the corresponding servant action.
Defining the action to be on an four dimensional manifold, $M$,  with boundary
$\partial M=\Sigma$, we obtain wavefunctionals by following the same
steps we took in the three dimensional case.
This results in
a functional $\Psi [e|_{\Sigma}]$.  The proof that the wavefunctional
so constructed satisfies the constraints of quantum gravity is now
the same as that of Hartle-Hawking \cite{HarHawk}.
We differ the exact results, in particular, a presentation of the
normalization procedure,  to a  future work \cite{BroLif}.

\vskip 0.5truein
\section{Conclusions}
\medskip\forcepar
In this work we have indicated a possible way of using TQFT's
 to construct
wavefunctionals and physical correlation functions in three and four
 dimensional quantum gravity. We gave explicit results in the three
 dimension case and laid the building blocks for the construction
in four dimensions.
Along the way we have shown that in quantum gravity,
there are functions of the fields
whose vacuum expectation values are not only diffeomorphism invariants of
spacetime, but also of geometrical significance on moduli space.
A possible definition of an inner product was mentioned and
will be elaborated in \cite{BroLif}.
This work also
indicates that it might be useful to consider quantum gravity
in a larger geometrical
setting than usual.

In concluding, it is tempting to speculate that pursuit along the
lines advocated in this work may lead to possible field theoretic
relations between intersection numbers on a manifold and the
intersection numbers on the moduli space of field configurations for
sections of bundles over
that manifold.  In particular, we know that the Wilson loop
observables in \bfts{} \cite{Wit(Jones)} and the loop observables
\cite{Smol}
construct knot invariants.  Well, we have found the projections of
Donaldson-like polynomial invariants into three and four dimensional
quantum gravity.  From a purely field theoretical point-of-view, we
then expect to find a relation between these two sets of operators.
\vskip 0.5truein
\centerline{\Large\bf\underline{Acknowledgements}}
\vskip 0.3truein
\forcepar
We thank M. Blau for a critical reading of the manuscript.
\renewcommand{\theequation}{\thesection.\arabic{equation}}
\newpage
\vskip 0.5truein
\appendix{}
\centerline{\Large\bf \underline{Appendices}}
\vskip 0.3truein
\section{3D BF Gauge Theory}\label{A_BF}
\setcounter{equation}{0}
\medskip
\forcepar
Recall \cite{TFT_rev} that the actions of \bfts{} are topological
and of the form
\be
S_{BF}\seq \int_M Tr(B\wedge F)\ \ ,\ee
where $B$ is an $ad(G)$-valued $(n-2)$-form on the oriented, closed
$n$-manifold, $M$ and $F$ is the curvature of the $G$-bundle whose
connection is $A$.  Path integrals for these theories with the
insertion of any operators except those composed of $B$ have support
only on flat connections.  The wavefunctionals for these theories reduce
to $L^2$-functions on the moduli space, $\cm$, of flat connections.
For purposes of path integral quantization, the partition function of
the \bft is
\be
Z_{BF}\seq \int[dA][dB][d{\bar c}][d c][d{\bar c}'][dc'][db][db']
e^{-[S_{BF}+S_{BF,gf}]}\ \ ,\ee
where
\be
S_{BF,gf}\seq \int_MTr(bd\dual A\pls  b' d_A\dual B\mi {\bar c}d\dual
d_A c\mi {\bar c}'d\dual d_A c')\ \ .\ee
The last action represents the projection of the connection into the
Lorentz gauge  and  the removal of the covariantly exact part of $B$;
all done by means of the symmetries of the \bft.  In this gauge fixing,
the $c(c')$ and ${\bar c}({\bar c}')$ fields are the zero-form,
anti-commuting ghosts  and anti-ghosts for the $A(B)$ projections,
respectively.

Canonical quantization on $M=\Sigma\times\IR$  immediately leads to
the constraints \cite{Hor},
\be
\dual d_A B~\approx~ 0\ \ ,\qquad \dual F~\approx~ 0\ \ ,\ee
where the Hodge dual here is defined on $\Sigma$ and is induced from
that on $M$.  The first of these constraints is Gauss's law
enforcing the  gauge invariance of physical states and the second
requires that these states have support only on flat connections.
In this special case of 2+1 dimensional quantum gravity, it can be
shown \cite{AshHus_etal,Wit(CSGrav)} that
on physical states $Diff(\Sigma)$ is equivalent to these constraints.

\bigskip
\section{3D Super-BF Gauge Theory}\label{A_SBF}
\setcounter{equation}{0}
\medskip
\forcepar
The action for \sbft{} \cite{TFT_rev,BlThom} is
\bea
S_{SBF}\seq \int_M\hskip -8pt &Tr&\hskip -8pt\{B\wedge F  \mi \chi
\wedge
d_A\psi\nen
&+&\eta
d_A\dual\psi \pls \lambda  \dual \triangle_A \phi
\pls \lambda [\psi,\dual\psi]\nen
&+&\eta'
d_A\dual \chi \pls \lambda' {} \dual \triangle_A \phi'
\pls \lambda' [\psi,\dual\chi]~\} \ \ .\label{E_SBF(action)}
\eea
All fields are $ad(G)$ valued and their
form degree, Grassmann-parity and fermion/ghost number are listed in
the following table:
\begin{center}
\begin{tabular}{|c|c|c|c|}\hline
FIELD&DEGREE&G-PARITY&GHOST \#\\ \hline\hline
$B$&$1$&even&~~0\\ \hline
$A$&$1$&even&~~0\\ \hline
$\chi$&$2$&odd&$-1$\\ \hline
$\psi$&$1$&odd&$~~1$\\ \hline
$\eta$&$0$&odd&$-1$\\ \hline
$\eta'$&$0$&odd&$~~1$\\ \hline
$\lambda$&$0$&even&$-2$\\ \hline
$\phi$&$0$&even&$~~2$\\ \hline
$\lambda'$&$0$&even&$~~0$\\ \hline
$\phi'$&$0$&even&$~~0$\\ \hline
\end{tabular}
\end{center}
Placing this set of fields in the context of section \ref{Sec_Idea},
the \bft{} is the GFT and \sbft{} is the TQFT.  The sets of fields are
represented by $X=(B,A)$ with $Y$ being the rest of the fields in this table.

The (Yang-Mills) gauge invariant action (\ref{E_SBF(action)}) may be
obtained by starting with the zero lagrangian and gauge fixing the
topological
symmetry $\ddelta A=\epsilon\psi$ to the flat connection condition,
$F=0$.  It is invariant under the horizontal  BRST
transformations
\bea
&&Q^H:A\to\psi\ \ ,\qquad \quad~~ Q^H:\psi\to d_A\phi \ \
,\nen
&&Q^H:\chi\to B\pls d_A\phi'\ \ ,\qquad Q^H:B\to
[\chi,\phi] \pls [\phi',\psi]\ \ ,\nen
&&Q^H:\lambda\to\eta \ \ ,\qquad\quad
Q^H:\eta \to [\lambda,\phi] \ \ ,\nen
&&Q^H:\lambda'\to\eta' \ \ ,\qquad\quad
Q^H:\eta' \to [\lambda',\phi] \ \ .\label{BRST}
\eea
Additionally,  it is
invariant under the one-form symmetry $\delta B= d_A
\Lambda$.  The gauge fixing of these symmetries introduces
the usual ghost ``kinetic" terms plus some new terms which
involve  Yukawa-like couplings with $\psi$ and $B$.   We
will return to these later. It is worthwhile to note that the action,
$S_{SBF}$ may be written as the action of a \bft{} plus ``supersymmetric"
completion term as\footnote{We use $S_S$ heavily in the text.}:
\be
S_{SBF}\seq S_{BF}\pls S_S\ \ .\ee

The partition function for
\sbft{} is (see appendix \ref{A_Nota} for our notation)
\be
Z(M)\seq \int [d\mu]_{SBF} e^{-S_{SBF}}\ \ .\ee
Integrating
over $B$ we see that this partition function has support
only on flat connections as is the case with the \bfts.
However, $\psi$ and $\phi$ have made their appearances.
The observables \cite{TFT_rev} of this theory are elements of
the $Q^H$-equivariant cohomology and are maps from  $H_*(M)$ to
$H^*(\cm)$.  For rank two groups they are constructed as polynomials of
the following homology cycle integrals:
\bea
\co^{(4)}&\seq& \half Tr(\phi^2)\ \ ,\nen
\co^{(3)}&\seq& \int_\gamma Tr(\phi\psi)\ \ ,\nen
\co^{(2)}&\seq& \int_\Gamma Tr(\phi F\pls\half\psi\wedge\psi)\ \ ,\nen
\co^{(1)}&\seq& \int_M Tr(\psi\wedge F)\ \ .\eea
In these expressions, the index $(k)$ represents
the fact that $\co^{(k)}$ is a $k$-form on moduli space or carries
ghost number $k$ in the BRST language.
These are the three dimensional analogs of the Donaldson invariants
which may be constructed in four dimensional topological
Yang-Mills theory \cite{Wit(TYM)}.
\newpage
\section{Special Topology for BF: $\Sigma_g\times S^1$}\label{A_Spec}
\medskip
\setcounter{equation}{0}
\forcepar
In this appendix, we would like to suggest the possible existence of polynomial
invariants  in the pure
\bft{} if the
three-dimensional manifold is taken to be the Lens space $S^2\times
S^1$ or $\Sigma_g\times S^1$, in general\footnote{
Here, as in the text, $\Sigma_g$ is a genus $g$ Riemann surface.}.
In what follows we will assume the temporal gauge.  However, this is not
completely possible due to the holonomy of the gauge field in the $S^1$
direction.  It is for this reason that our discussion is only suggestive.
Nevertheless, see ref. \cite{BT(CS)} in which an explicit demonstration of the
relation between Chern-Simons theory and G/G WZW theory on $\Sigma_g$ is given.

Expand all of the fields in the harmonics of $S^1$, $e^{in\theta}$,
(where $\theta$ is the coordinate on $S^1$ and $n$ is an integer)
symbolically as $\Phi(\Sigma_g\times S^1)\equiv
\sum_n\Phi_{(n)}(\Sigma_g) e^{in\theta}$.  Then choose the ``temporal
gauge'' so that the  connection in the $S^1$ direction vanishes
leading to the action\footnote{In this appendix, $\phi$ is not the
scalar field in the \sbft{} considered in the body of the paper.}
\bea
S_{BF}\seq \int_{\Sigma_g}Tr\bigg(&&\sum_n \phi_{(n)} d A_{(-n)} \pls
\sum_{n,m} \phi_{(n)} A_{(m)}\wedge A_{(-n-m)}\nen
&&\mi i \sum_n n B_{(n)}A_{(-n)} \pls \sum_n n{\bar c}_{(n)}
c_{(-n)}~\bigg)\ \ ,\label{S_SIG}\eea
where $\phi_{(n)}$ are the components of the original $B$ field in
the $S^1$ direction.
Realizing that the $B_{(n)}$ for $n\neq0$ does not appear in a term
with derivatives, we integrate it out of the action finding
\be
S_{BF}\seq \int_{\Sigma_g}Tr\bigg(\phi F\pls \sum_n n \bar c_{(n)}
c_{(-n)}\bigg)\ \ ,\label{S_Sig}\ee
where $\phi\equiv \phi_{(0)}$ and $F$ is the curvature on $\Sigma_g$
constructed out of $A_{(0)}$.  With the exception of the completely
decoupled fermionic term, this is the action for two dimensional $BF$
theory.  It has been recently studied quite extensively
\cite{Wit(2DYM),Thom}.  Notice that we did not obtain it via
compactifying the $S^1$ direction.  That direction simply decouples
due to the first order and off-diagonal nature of the gauge theory.
Extending the theory to incorporate an equivariant
cohomology is possible, however, we will not need this in order to
obtain our results.

{}From the partition function for this action, it is easy to show that
gauge invariant functions of $\phi$ will be invariant under
diffeomorphisms.  As ${\cal L}_K\phi=i_kd_A\phi + [\phi,\alpha(K)]$,
we must show that $\ev{d\co(\phi)}=0$, where $\co$ is some gauge
invariant function constructed only out of $\phi$.  This equality
follows from the statement that ${d_A\phi}$ is obtained by varying
the action with respect to the connection, thus its expectation value
and/or correlation function with any other functions of $\phi$ is a
total functional derivative on $\ca$; hence it vanishes.  Another way
to see this result is that a symmetry of the action (\ref{S_SIG})
exists in which $\phi$ may be shifted into a ${\bar c}_n$ (or $c_n$).
This symmetry, however, does not affect the $\phi$ zero-mode
(solution of $d_A\phi=0$) as it does not appear in the action.
The exponent in definition of $\cb_I$ is $Q^H$ exact and since it is
constructed to be gauge invariant it is also $Q^H$ closed.

\vskip 0.5truein
\section{Notation}\label{A_Nota}
\setcounter{equation}{0}
\medskip
\forcepar
The symbol, $G$ is used to denote a semi-simple Lie group.
The space of gauge connections is written as $\ca$.
Our generic notation for spacetime manifolds is $M$ while  we
use the symbol, $\cm(M,G)$ for the moduli space of specific (e.g., flat)
connections for the $G$-bundle, $P$, over $M$.
The exterior derivative on $M$
is $d$ while the covariant exterior derivative with respect to the connection
$A$ is $d_A$.  Coordinates are $M$ are written as $x$, $y$, etc. while
coordinates on $\cm(M,G)$ denoted by $m$.  The gauge covariant
laplacians on $k$-forms are written as $\triangle_A^{(k)}$.  The
genus of a handlebody/Riemann surface is $g$.  Any metrics which appear
explicitly are written with indices or otherwise in an obvious manner.
While $\gamma$ denotes a homology one-cycle,
$\Gamma$ stands for a homology two-cycle.  The functional measures
used are defined in the
following table:
\begin{center}
\begin{tabular}{|c|c|c|}\hline
NOTATION&MEASURE&ACTION\\ \hline\hline
$[d\mu]_{BF}$&$[dA][dB][d{\bar c}][dc][d{\bar c}'][dc'][db][db']
$&$S_{BF} +S_{BF,gf}$\\ \hline
$[d\mu]_S$&$[d\chi][d\psi][d\eta][d\eta'][d\phi][d\lambda][d\phi'][d\lambda']$
&$S_S$\\ \hline
$[d\mu]_{SBF}$&$[d\mu]_{BF}[d\mu]_S$&$S_{SBF}$\\ \hline
\end{tabular}
\end{center}
All other notations are established in the text.  Except note that
our field notations for the TQFT's are not the same in three and
four dimensions (see section \ref{Sec_4D}).

\end{document}